\documentclass[aps,prl,reprint,floatfix,citeautoscript,longbibliography,superscriptaddress]{revtex4-1}
\usepackage{graphicx,float}
\usepackage{amsmath,amssymb,mathrsfs,dcolumn}
\usepackage{gensymb}
\usepackage[usenames]{color}
\usepackage{subfigure}
\usepackage{ulem}
\usepackage{hyperref}
\hypersetup{colorlinks=true, linkcolor=blue, citecolor=red, urlcolor=blue}
\begin{document}

\title{Magnon antibunching in a nanomagnet}
\author{H. Y. Yuan}
\affiliation{Institute for Theoretical Physics, Utrecht University, 3584CC Utrecht, The Netherlands}
\author{Rembert A. Duine}
\affiliation{Institute for Theoretical Physics, Utrecht University, 3584CC Utrecht, The Netherlands}
\affiliation{Center for Quantum Spintronics, Department of Physics, Norwegian University of Science and Technology, NO-7491 Trondheim, Norway}
\affiliation{Department of Applied Physics, Eindhoven University of Technology, P.O. Box 513, 5600 MB Eindhoven, The Netherlands}

\date{\today}

\begin{abstract}
We investigate the correlations of magnons inside a nanomagnet and identify a regime of parameters where the magnons
become antibunched, i.e., where there is a large probability for occupation of the single-magnon state.
This antibunched state is very different from magnons at thermal equilibrium and microwave-driven coherent magnons. We further obtain the steady state analytically and describe the magnon dynamics numerically, and ascertain the stability of such antibunched magnons over a large window of magnetic anisotropy, damping and temperature. This means that the antibunched magnon state is feasible in a wide class of low-damping magnetic nanoparticles. To detect this quantum effect, we propose to transfer the quantum information of magnons to photons by magnon-photon coupling and then measure the correlations of photons to retrieve the magnon correlations. Our findings may provide a promising platform to study quantum-classical transitions and for designing a single magnon source.
\end{abstract}

\maketitle

{\it Introduction.---} Magnon spintronics is a rising field with the goal to manipulate the collective excitations in ordered magnets, so-called magnons, as information carriers and to employ their advantage of long spin diffusion length, low energy consumption, and their integration capability with traditional electronics devices \cite{Chumak2015}. In the absence of external stimulation, the magnons reach thermal equilibrium with the environment, and form a thermal magnon gas, which influences the spontaneous magnetization, internal energy and specific heat of a macroscopic magnet.  
In modern spintronics, much attention is devoted to the generation, transport and read-out of nonequilibrium coherent magnons in magnetic layered structures under the assistance of external knobs from microwaves, terahertz waves, electric current, and temperature gradient  \cite{Kamp2010,Bauer2012,Nakayama2013,Lebrun2018}.
Magnonic phenomena  may be roughly classified into three classes: (i) pure magnon transport such as spin pumping \cite{Yaro2002}, spin Seebeck effect \cite{Uchida2008,Xiao2010}, magnon spin torque \cite{PYan2012}, (ii) interconversion between electrons and magnons including (inverse) spin Hall effect \cite{Sinova2015}, and spin Hall magnetoresistance \cite{Weiler2012,Chen2013}, and (iii) coherent or dissipative coupling of magnons with other quasi-particles including photons and phonons \cite{Soykal2010,Huebl2013, Cao2015, Vahram2018,Harder2018,Berk2019}.

With the rise of quantum information science and the tendency to dock magnonics with quantum information, it is becoming particularly important to ask whether quantum states of magnons such as Fock states, squeezed states, antibunching, and Schr\"{o}dinger cat states can be achieved in magnonic systems. Both the spintronics and quantum optics community have made a few steps towards this direction \cite{yuan2020afm,Kamra2019,Elyasi2020,Dany2020}. For example, the magnons in antiferromagnets have been shown to be a two-mode squeezed state with strong entanglement, while this entanglement is even enhanced by cavity photons through the cooling effect \cite{yuan2020afm,Kamra2019}. Moreover, a single magnon excitation was recently detected by entangling the magnons with a superconducting qubit sensor \cite{Dany2020}.

 \begin{figure}[H]
	\centering
	\includegraphics[width=0.5 \textwidth]{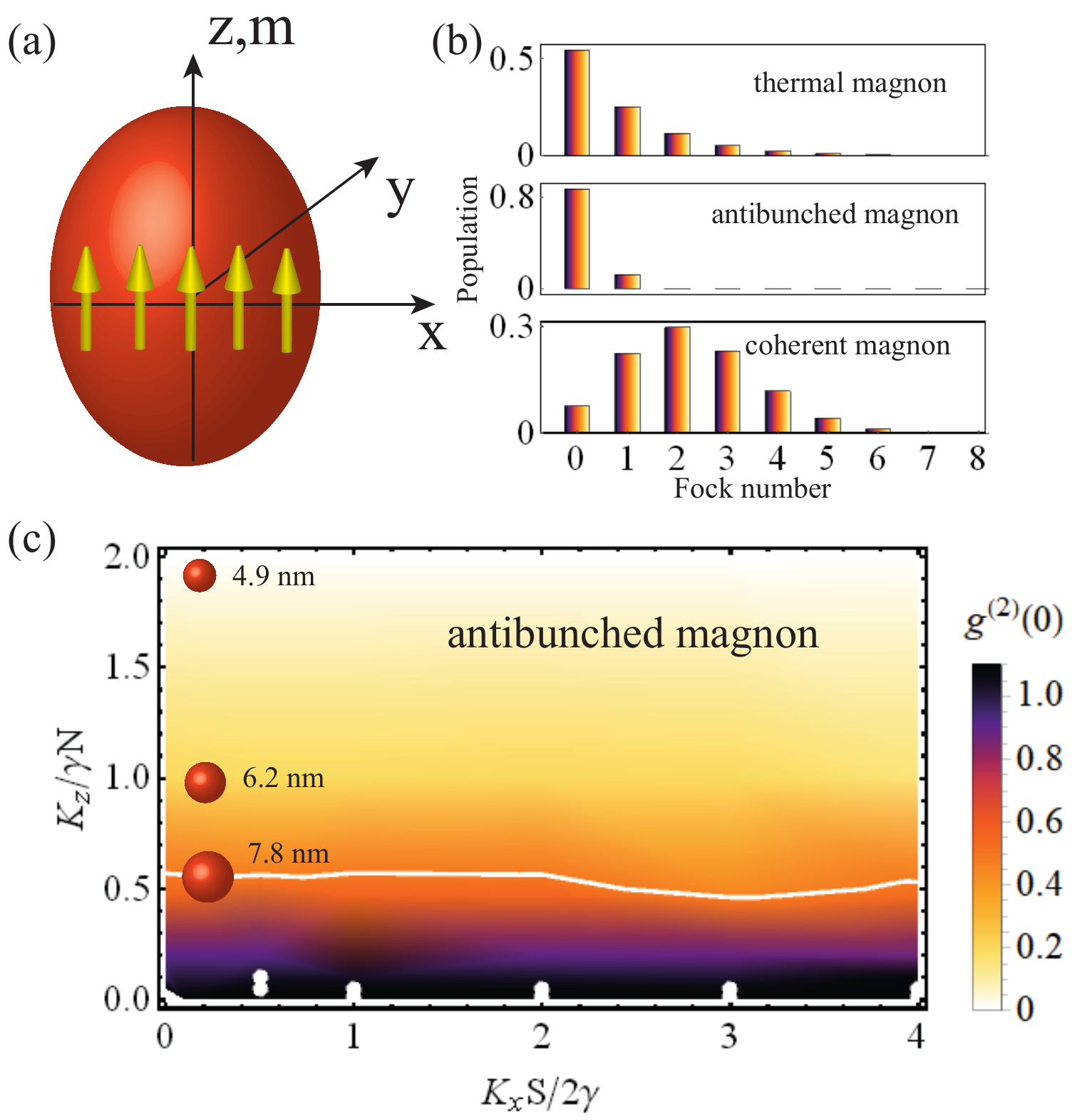}\\
	\caption{(a) Schematic of a nanomagnetic particle magnetized along the $z$ axis while the $x$ axis is a hard axis. (b) Fock distribution of the thermal, antibunched and coherent magnons, respectively.  (c) Phase diagram of the nanoparticle in the $(K_xS/2\gamma,K_z/\gamma N)$ plane. The white line corresponds to the second-order auto-correlation function $g^{(2)}(0)=0.5$. The size of the particle is calculated by assuming a lattice constant 0.5 nm. Other parameters are $\gamma=10^{-3}\omega_a, \xi=2\times 10^{-4}\omega_a,\omega=\omega_a+v,\omega_a/k_BT=9.21$.}
	\label{fig1}
\end{figure}

In this letter, we study magnon correlations inside a nanomagnet at low temperature analytically and numerically and find that the magnons become antibunched under the influence of magnon-magnon interactions. The single magnon excitation dominates the excitation spectrum, differing from the thermal magnon gas and the coherent magnons.
Antibunching becomes more pronounced in a magnet with small size of several nanometers, low magnetic damping on the order of $10^{-3}$ and environment temperature 0.2 Kelvin, which should be accessible in current experiments. Our findings may open an intriguing window to study the quantum properties of magnons and further benefit their docking with photons for quantum information processing.

{\it Model and method.---} We consider a biaxial nanomagnet as shown in Fig. \ref{fig1}(a), described by the Hamiltonian,
\begin{widetext}
\begin{equation}
\mathcal{H}=-J\sum_{\langle ij\rangle} \mathbf{S}_i\cdot \mathbf{S}_j -K_z \sum_i (S_i^z)^2 + K_x\sum_i (S_i^x)^2-H\sum_i S_i^z,
\end{equation}
\end{widetext}
where $\mathbf{S}_i$ is the spin vector of $i$th site with magnitude $S$, $J,K_z,K_x$ and $H$ are respectively the exchange coefficient, easy-axis and hard-axis anisotropy coefficients and external field. After a Holstein-Primakoff transformation \cite{HP1940} to third order in $a_i$, i.e., $S_i^+ =\sqrt{2S - a_i^\dagger a_i} a_i\approx \sqrt{2S}(a_i-a_i^\dagger a_ia_i/4S), S_i^- =a_i^\dagger\sqrt{2S - a_i^\dagger a_i} \approx \sqrt{2S}(a_i^\dagger-a_i^\dagger a_i^\dagger a_i/4S),S_i^z=S-a_i^\dagger a_i$, where $a_i$($a_i^\dagger$) is the magnon annihilation (creation) operators that satisfy the commutation relations $[a_i,a_j^\dagger]=\delta_{ij}$, where $S_i^\pm=S_i^x \pm iS_i^y$ are spin rising and lowering operators, the Hamiltonian can be written in momentum space up to fourth-order terms as,

\begin{widetext}
	\begin{equation}
	\mathcal{H}=\omega_a a^\dagger a + w(a^\dagger a^\dagger + aa) + v(a^\dagger a)^2+u(a^\dagger aaa +a^\dagger a^\dagger a^\dagger a) +
	\xi (a e^{i\omega t} + a^\dagger e^{-i\omega t})
	\label{Hamk}
	\end{equation}
\end{widetext}
where $a=a_{k\rightarrow 0}$ with $k$ the wavevector of magnons. At low temperature, the low energy magnons ($k\rightarrow0$) dominate the excitation spectrum, such that we neglect the contribution from high energy magnons \cite{notek}. Furthermore, $\omega_a=2K_z+K_x+H$, $w=K_xS/2$, $v=-K_z/N-K_x/2N$, $u =-K_x/4N$, with $N$ the number of spins in the magnet. The last linear term is added as a driving from microwave with strength $\xi$ and frequency $\omega$.

The quantum correlations among magnons are characterized by the zero-delay second-order autocorrelation function \cite{DFWalls},

\begin{equation}
g^{(2)}(0)=\frac{\langle a^\dagger a^\dagger a a \rangle}{\langle a^\dagger a \rangle^2},
\end{equation}
where $\langle A \rangle=tr(\rho A)$ is the ensemble average of the observable $A$ with $\rho$ the density matrix of the system. In general, $g^{(2)}(0)>1$
refers to a classical correlation of magnons, for example $g^{(2)}(0)=2$ for a thermal equilibrium magnon gas. The critical case $g^{(2)}(0)=1$ corresponds to coherent magnons. $g^{(2)}(0)<1$ means the magnons are antibunched, a purely quantum-mechanical type of behavior with Fock number distribution as shown in Fig. \ref{fig1}(b). Moreover, $g^{(2)}(0)=0$ indicates that a perfect single magnon source is realized. To calculate the correlations of magnons, we numerically solve the Lindblad master equation for the time evolution of the density matrix of the system \cite{Lindblad1976},

\begin{equation}
\frac{\partial \rho}{\partial t}=-i[\rho, \mathcal{H}] +\mathcal{L}\rho,
\end{equation}
where $\mathcal{L}\rho= \sum_{n=1,2} [C_n\rho C_n^\dagger-(\rho C_n^\dagger C_n+ C_n^\dagger C_n\rho)/2]$, with
 $C_1=\sqrt{(n_\mathrm{th}+1)\gamma} a$ and $C_2=\sqrt{\gamma n_\mathrm{th}} a^\dagger$ that describe magnon annihilation and creation, respectively, $\gamma$ is the dissipation strength which is related to the dimensionless Gilbert damping $\alpha$ via $\gamma=\alpha \omega_a$\cite{notedamp}, $n_\mathrm{th}$ is the magnon population in thermal equilibrium.

{\it Phase diagram.---}The full phase diagram in the $(K_xS/2\gamma,K_z/\gamma N)$ plane is shown in Fig. \ref{fig1}(c). (i) For a magnet with rotational symmetry around the $z$ axis ($K_x=0$), the magnons are able to reach an antibunched steady state.  The smaller the size of the magnet ($N$), the smaller $g^{(2)}(0)$ and thus the magnons antibunching behavior becomes more pronounced. (ii) Once the rotational symmetry is broken ($K_x \neq 0$), the populations of magnons will keep oscillating but the average correlation function $g^{(2)}(0)$ is still below one. (iii) A quantum-classical oscillation is found for weak easy-axis anisotropy, as indicated by the white dots. Next, we will study the essential physics of these regimes in detail.

{\it Steady antibunching.---} In the absence of the hard-axis anisotropy, $w=u=0$, the Hamiltonian can be rewritten in a rotating frame $V=\exp(-i\omega a^\dagger a t )$ as,
\begin{equation}
\mathcal{H}=\Delta_a a^\dagger a + v (a^\dagger a)^2 +\xi (a^\dagger + a),
\end{equation}
where $\Delta_a$ is the detuning between microwave driving and magnon frequency defined as $\Delta_a \equiv \omega_a-\omega-i\gamma/2$ with dissipation strength included as $\gamma$.

At low temperature, thermal mixture of magnons at different Fock states is expected to be very weak, such that the density matrix may be approximated as a pure state $\rho=|\varphi \rangle \langle \varphi |$, where $|\varphi \rangle =\sum_{n=0}^\infty C_n|n \rangle$ and its evolution satisfies the Schr\"{o}dinger equation $i\partial_t |\varphi \rangle = \mathcal{H} |\varphi \rangle$. The steady state ($\partial C_n/\partial t=0$) implies the recursion relation,
\begin{equation}
(\Delta_a n + v n^2)C_n + \xi \sqrt{n+1}C_{n+1} + \xi \sqrt{n}C_{n-1}=0.
\end{equation}
Because the magnon density is very small, the ground state population dominates the system ($C_0\approx 1$). By making a cut-off at $n=3$, we are able to analytically solve for the probability amplitude,
\begin{equation}
\left (\begin{array}{c}
C_1 \\C_2
\end{array} \right )=
\frac{1}{(\Delta_a + v)(\Delta_a + 2v)-2\xi^2}
\left (\begin{array}{c}
-(\Delta_a + 2v)\xi \\ \xi^2/\sqrt{2}.
\end{array} \right ).
\end{equation}
This yields for the second-order correlation function of magnons,
\begin{equation}
g^{(2)} (0)=\frac{\langle a^\dagger a^\dagger a a\rangle}{\langle a^\dagger a\rangle ^2}=
\frac{|\Delta_a (\Delta_a +v)-\xi^2|^2}{(|\Delta_a+v|^2+\xi^2)^2}.
\label{g2steady}
\end{equation}

\begin{figure}
	\centering
	\includegraphics[width=0.5\textwidth]{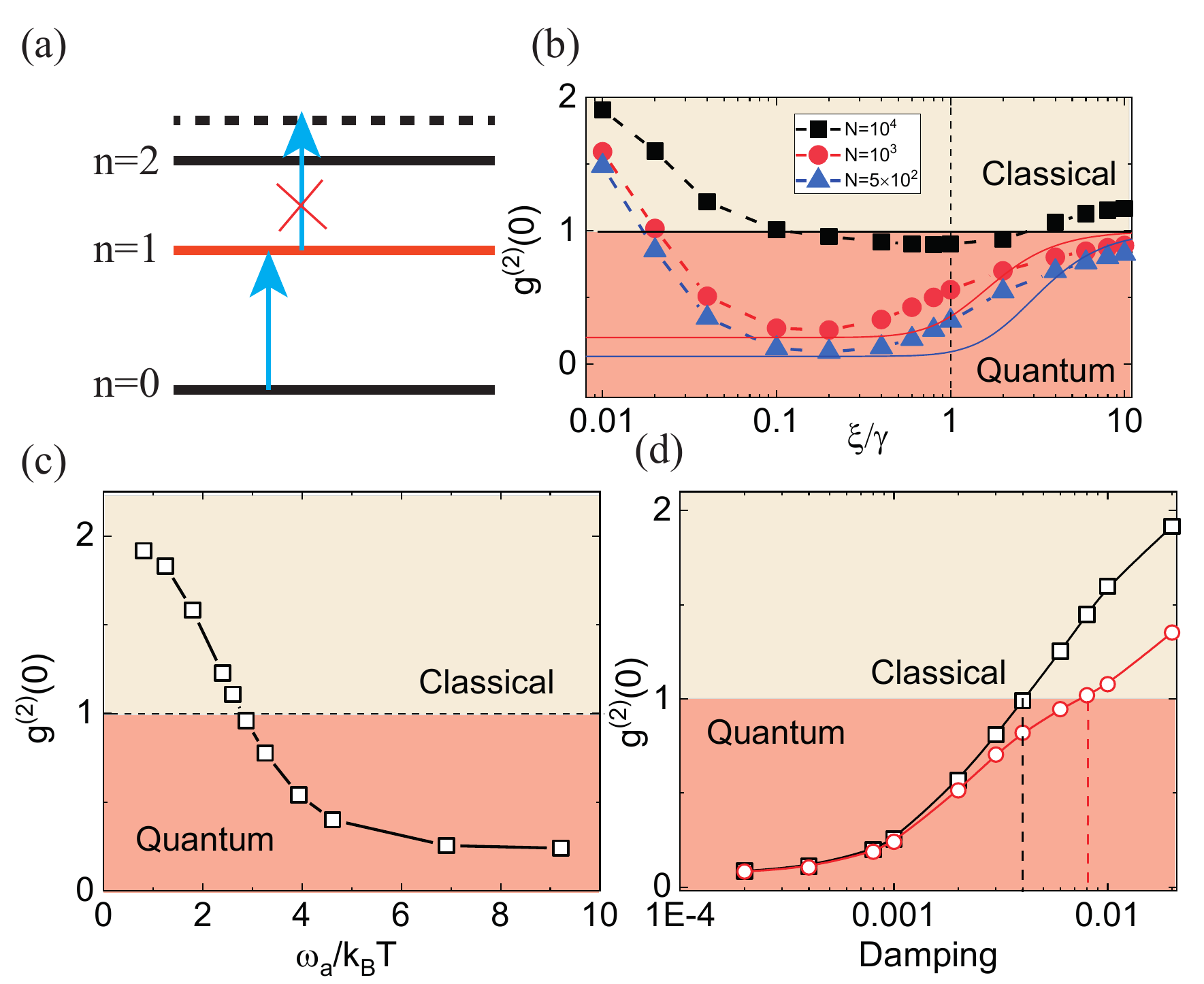}\\
	\caption{(a) Physical picture of magnon excitation under the influence of nonlinearity. (b) Driving dependence of the correlation function $g^{(2)}(0)$ for $N=10^4$ (black squares), $10^3$ (red circles), $5\times 10^2$ (blue triangles), respectively. The solid lines are
		theoretical predictions at zero temperature and the dashed lines are to guide the eyes. $n_{\mathrm{th}}=10^{-3},\gamma=10^{-3}\omega_a$. (c) Temperature dependence of the correlation function. $\xi/\gamma=0.2$. (d) Damping dependence of the correlation function at $n_{\mathrm{th}}=10^{-4}$ (red circles) and $10^{-3}$ (black squares).}
	\label{fig2}
\end{figure}

When the driving is largely detuned $|\omega_a-\omega| \gg \gamma, v$, the leading contribution to the correlation is equal to,

\begin{equation}
g^{(2)} (0)\approx \frac{1}{(1+v/(\omega_a-\omega))^2},
\end{equation}
indicating that no apparent quantum correlation of magnons is expected, especially when the magnon-magnon interaction is absent ($v=0$). Of particular interest is when the driving field reaches resonance with the first excitation level ($\omega=E_1-E_0=\omega_a+v$), Eq. (\ref{g2steady}) is reduced to

\begin{equation}
g^{(2)} (0)=\frac{(1/4+\xi^2/\gamma^2)^2 +  v^2/4\gamma^2}{(1/4+\xi^2/\gamma^2 + v^2/\gamma^2)^2}.
\end{equation}
If the driving is sufficiently strong $\xi \gg \gamma,v$, we have $g^{(2)} (0)= 1$. In this regime, we have that $\langle (\Delta x)^2 \rangle +\langle (\Delta p)^2 \rangle =1/2$, where $x=(a^\dagger+a)/2$ and $p=i(a^\dagger + a)/2$ are the quadrature components of the magnon mode. This minimum uncertainty relation implies that the steady state is a coherent state, which is expected for strong driving. On other hand, when $v\gg \gamma \gg \xi$, $g^{(2)} (0)= \gamma^2/v^2 \ll 1$, which suggests that the magnons become antibunched. The essential physics can be understood from Fig. \ref{fig2}(a). The energy levels of the system without driving are $E_n=(n \omega_a + n^2 v), (n=0,1,2,...)$, which has a nonlinear dependence on $n$. The input microwave with $\omega=\omega_a +v$ will efficiently excite the magnons to level $n=1$, but cannot excite them further due to the frequency mismatch. Hence, the population of magnons at $n=1$ becomes significant compared to other excited states and then results in a single-magnon-like state with  $g^{(2)} (0) <1$.

When thermal effects become significant, the approximation of the density matrix as a pure state is not valid any longer. Figure \ref{fig2}(b) shows the driving dependence of the
correlation function at a finite temperature for various sizes of magnet, found from solving the Lindblad master equation.
It is consistent with the expectation that the magnons evolve to a coherent state under strong driving ($\xi$) and the magnons becomes strongly antibunched for a system with stronger nonlinearity ($v$) or smaller spin number $N$. The difference is at weak driving, where the real correlation function should be equal to 2 instead of 1, predicted by the theory. This is because the thermal magnons dominate the systems in this regime and thus the full density matrix needs to be considered instead of a pure state in a reduced Hilbert space.

To further see the crossover between classical magnons and quantum magnons caused by the temperature, we tune the temperature and show the results in Fig. \ref{fig2}(c). The typical transition temperature is around $\omega_a/k_BT=3$, corresponding to 0.2 Kelvin for magnetic resonant frequency $\omega_a/2\pi=10$ GHz. Such low temperature is sufficient to guarantee the stability of the magnetic particle because of $K_zV/k_BT \gg 45$ \cite{Dieny2017}. The damping dependence of the correlation function at different temperatures is shown in Fig. \ref{fig2}(d). The lower the temperature, the larger the magnetic damping ($\gamma/\omega_a$) is allowed to observe the quantum behavior of magnons. The typical value of damping to observe antibunching is around 0.008, which is a realistic value for many magnetic nanoparticles, such as Cobalt, permalloy.

\begin{figure}
	\centering
	\includegraphics[width=0.5 \textwidth]{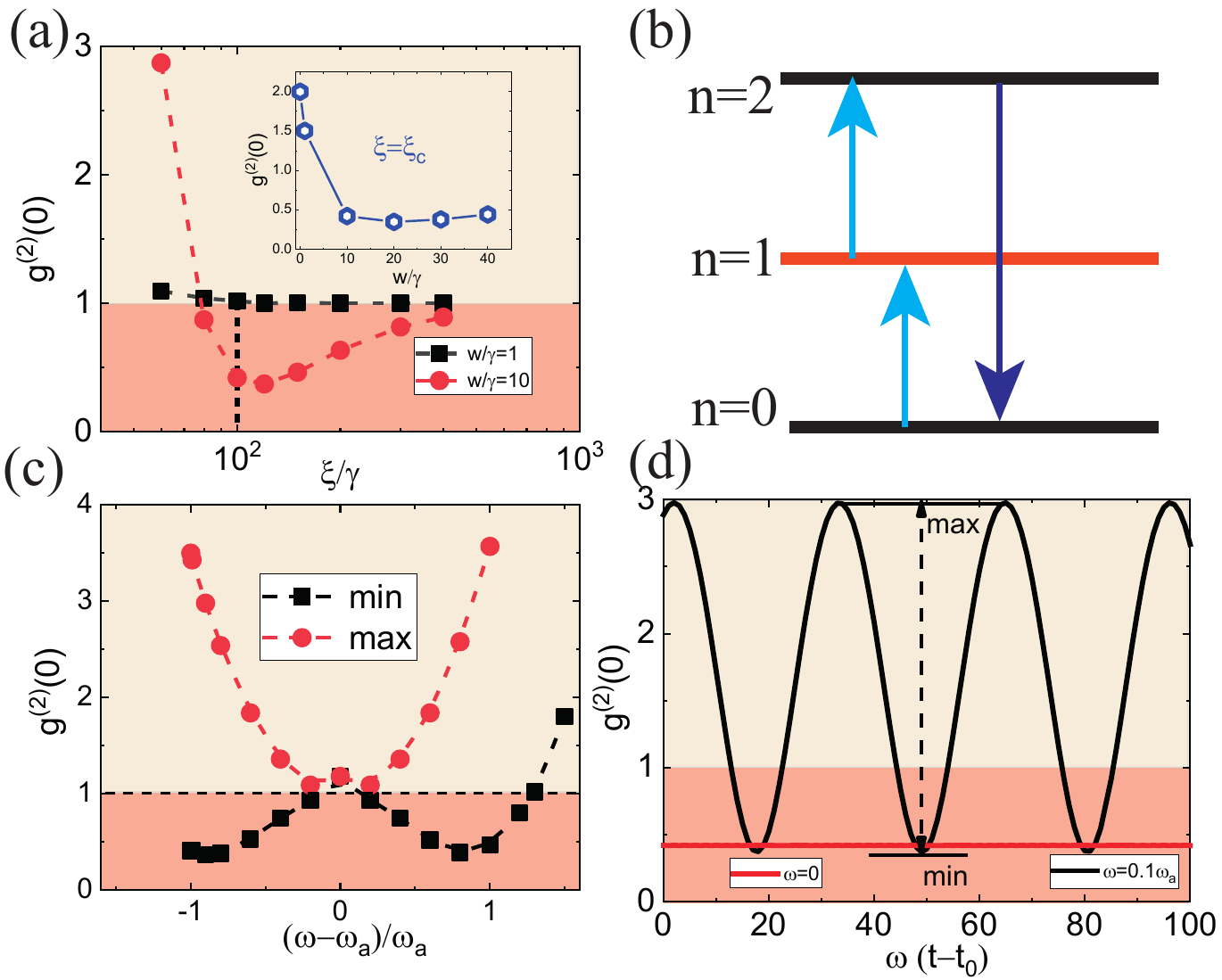}\\
	\caption{ (a) Driving dependence of the correlation function $g^{(2)}(0)$ for $w/\gamma=1$  (black squares), and $10$ (red circles), respectively. The inset shows the correlation as a function of $w/\gamma$ when $\xi=\sqrt{\omega_a w}$. (b) Physical picture of magnon excitation under the influence of nonlinearity. (c) Detuning dependence of the correlation function. $n_{\mathrm{th}}=10^{-3},\gamma=10^{-3}\omega_a,w=0.01\omega_a,\xi=0.1\omega_a$. (d) Time evolution of correlation function at $\omega=0$ (red line) and $0.1\omega_a$ (black line), respectively.}
	\label{fig3}
\end{figure}

{\it Oscillating antibunching.---} The physics will be quite different in an easy-plane  magnet ($K_z=0,K_x>0$), now the Hamiltonian becomes
\begin{equation}
	\mathcal{H}=\omega_a a^\dagger a + w (a^\dagger a^\dagger + aa) +\xi (a^\dagger e^{-i\omega t} + ae^{i\omega t}),
\label{osHam}
\end{equation}
where the higher-order terms are omitted for their smallness compared with the two-magnon process, that we shall see, already leads to bunching. In the limit that the field is time-independent ($\omega=0$) \cite{note01}, we can follow a similar analytical treatment as in the previous section and derive,

\begin{equation}
\left (\begin{array}{c}
C_1 \\C_2
\end{array} \right )=
\frac{1}{\Delta_a^2-\xi^2}
\left (\begin{array}{c}
-(\Delta_a -w)\xi \\ (\xi^2-w\Delta_a)/\sqrt{2}
\end{array} \right ).
\end{equation}
One immediately sees that there exists a special point, $\xi_c=\sqrt{\omega_a w}$, at which the occupation probability of level $n=2$ equals to zero due to the smallness of $\gamma$. This indicates the existence of a perfect antibunching, which is confirmed by finite temperature simulations shown in Fig. \ref{fig3}(a).
The essential physics can be understood in Fig. \ref{fig3}(b) as follows. Now the linear driving term tends to drive the magnon from $n-1$ to $n$, while this process competes with the double magnon decay caused by the parametric-type interaction $w$. At a certain point $\xi_c$, these two processes compensate and the net occupation of the $n=2$ is again very small. Once the driving increases further above this critical value, the coherent nature of the system becomes significant and thus $g^{(2)}(0) \rightarrow 1$, as shown in Fig. \ref{fig3}(a).

Under a time-dependent driving, this fine balance between the magnon driving and decay will not always happen, and thus the correlation function of the magnons oscillates with time. We confirm this prediction in Fig. \ref{fig3}(c) and \ref{fig3}(d), where the magnons keep oscillating between an antibunched state and classical state.


{\it Experimental relevance.---} As shown above, the essential conditions to produce antibunched magnons is to use a low-damping nano-sized magnet at low temperatures. Many reported magnetic materials such as $\mathrm{Co}$, $\mathrm{Ni_{20}Fe_{80}}$, $\mathrm{Mn_3Ga}$, $\mathrm{Fe_2O_3}$ and the two dimensional magnet $\mathrm{CrI_3}$ meet these conditions \cite{Thirion2003,Dieny2017,Oyarzun2015,Zhao2015}.


To probe the antibunched magnons, we suggest coupling the nanomagnet with a microwave, now the total Hamiltonian in the rotating frame becomes,

\begin{equation}
\begin{aligned}
\mathcal{H}_t&=\mathcal{H} + \Delta_c c^\dagger c  +g_{\rm mp}(a^\dagger c + a c^\dagger),
\end{aligned}
\end{equation}
where $\Delta_c = \omega_c -\omega -i \gamma_c/2$, $c (c^\dagger)$ is photon annihilation(creation) operator, $\omega_c, \gamma_c$ are photon frequency and dissipation, $g_{\rm mp}$ is the coupling strength between the magnon and photon. Note that $g_{\rm mp}=g_s\sqrt{N}$ depends on the resonator type and materials, and $g_s/2\pi$ can reach 10 KHz using a proper setup \cite{Hou2019}. For $N=10^3$ required for magnon antibunching, the hybrid system is typically in the weak coupling regime, i.e., $g_{\rm mp} < \gamma,\gamma_c,v$.

To obtain the steady state, we consider the evolution in the reduced Fock space as
$|\varphi \rangle =C_0|0 0\rangle+C_1|01 \rangle + C_2|10 \rangle + C_3|11 \rangle+ C_4|02 \rangle+C_5|20 \rangle$, where $|mn\rangle$ refers to the occupations of magnons ($m$) and photons ($n$), respectively, and the coefficients in steady state are computed by assuming $C_0 \approx 1$, from which the correlation function is obtained. Owing to the small coupling, we expand the correlation to the linear order of $g_{\rm mp}$ and then derive the ratio of magnon and photon correlation functions, where the later is denoted as $g_c^{(2)} (0)$,

\begin{equation}
\varrho=\frac{g_c^{(2)} (0)}{g^{(2)} (0)}=\left | 1+\frac{v}{v+\Delta_a + \Delta_c} \right |^2.
\end{equation}
As long as we choose a cavity with dissipation larger than the non-linear term, i.e., $\gamma_c \gg v$, the ratio $\rho$ is almost always equal to one, regardless of the magnitude of detuning $\omega_{a,c}-\omega$. This implies that the magnon correlation can be completely encoded into the cavity photons, and thus be measured by a Hanbury Brown-Twiss interferometer \cite{HBT}.

{\it Discussions and Conclusion.---}
We note that the magnon antibunching was also theoretically studied in hybrid magnet-cavity-superconductor qubit system \cite{Liu2019,Xie2020}, where the superconducting qubit induces an anharmonic energy levels of the system and thus suppresses the two magnon excitation, and magnet-cavity system with balanced gain and loss \cite{Wang2020}.
The cavity photon plays a crucial bridging role in these proposals, while it is not essential in the generation of antibunching that is discussed here.

In conclusion, we have found a classical-quantum transition in a nanomagnet as we reduce the temperature. In the quantum regime, the antibunched magnons prevail, with very different statistics as compared with thermal magnons and coherent magnons. The underlying physics is well understood as a result of magnon-magnon interactions, which either produce a nonlinear energy level distribution or enhance the multi-magnon process in the system. Our findings may open up an interesting perspective to create single photons using magnonic systems and further benefit the integration of magnons with photons to achieve quantum information transfer and quantum communications.

{\it Acknowledgements.---} This project has received funding from the European Research Council (ERC) under the European Union’s Horizon 2020 research and innovation programme (grant agreement No. 725509). RD is member of the D-ITP consortium, a program of the Netherlands Organisation for Scientific
Research (NWO) that is funded by the Dutch Ministry of Education, Culture and Science (OCW).

\clearpage
\onecolumngrid

\begin{thebibliography}{}
\bibitem{notek} We have neglected the interactions between the magnon modes at $k=0$ and $k\neq0$; this is justified in a small magnetic nanoparticles below the exchange limit as follows. The first energy level of non-uniform spin-wave is $\omega=2JSd^2k^2=2JSN^{-2/3}= 3~\mathrm{ THz} \gg \omega_a$, where $d=0.5 ~\mathrm{nm}$ is the lattice constant, $k=2 \pi N^{-1/3}/d$ is the wavevector of spin wave. This implies that the non-uniform spin wave is barely excited as compared with the uniform mode, and thus it is reasonable to disregard the magnons with $k\neq0$.
\bibitem{notedamp} Here, we assume for simplicity that the dissipation $\gamma$ is independent of energy and ignore effects of interactions on the dissipation.
\bibitem{note01} This limiting case corresponds to the application of a static field in the $x$ axis of our model. Strictly speaking, the ground state will deviate from the $z$ axis and thus the Holstein-Primakoff transformation around $z$ axis is no longer reliable. Here we only consider this case to understand the essential physics when $\omega \neq 0$. Furthermore, the model Eq. (\ref{osHam}) is meaningful when we consider a magnet under parametric pumping, which will be discussed in detail elsewhere.
\end{thebibliography}
\end{document}